# Application des techniques de compilation classique pour l'analyse syntaxique et sémantique des spécifications écrites en Object Z

**Informatique/génie logiciel**

**Fethi FKI[*,**], Kais HADDAR[**]**

*Faculté des sciences d'économie et de gestion de Mahdia*

*Sidi Messaoud 5111 Mahdia, Tunisie*

**Faculté des sciences de Sfax*

*B.P.809, 3018 Sfax, Tunisie*

*fethifki@yahoo.fr, Kais.haddar@fss.rnu.tn*

**RÉSUMÉ.** La construction d'un analyseur syntaxique pour un langage de spécification formelle tel que l'Object Z n'est pas une tâche facile. En effet, elle nécessite une double compétence aussi bien dans le domaine de la compilation que dans le domaine de la spécification formelle. Dans cet article, nous présentons tout d'abord quelques outils de vérifications des spécifications écrites en Z et Object Z en montrant les caractéristiques de chacun d'eux. Ensuite, nous identifions quelques contraintes sémantiques fréquentes en Object Z. Finalement, nous proposons une démarche pour la construction d'un analyseur syntaxique Object Z basé sur les techniques classiques de la compilation.

**MOTS-CLÉS:** Object Z, analyse syntaxique, contraintes sémantiques, techniques de compilation.

# 1. Introduction

Le code source d'un programme écrit dans un langage donné doit être compilé afin d'être transcris en langage machine, le seul qui soit compréhensible par les systèmes informatiques. On obtient alors un fichier exécutable [1]. Cette définition est valide pour les compilateurs des langages de programmation qui traduisent un texte source écrit dans un langage de haut niveau vers un langage utilisant des commandes exploitables par une machine informatique, ce qui n'est pas le cas pour les compilateurs des spécifications formelles. Pour ce type de langage, on ne peut pas parler d'une compilation dans le sens d'une traduction d'un texte dans un langage machine. Par conséquent, on ne peut pas parler d'un exécutable. Cela peut être expliqué par la nature conceptuelle des langages de spécification formelle, qui se caractérisent par un grand niveau d'abstraction, qui s'échappe des contraintes matérielles imposées par les différentes plates formes. La solution de ce problème était, pour les développeurs informatiques, la réalisation des vérificateurs de types et des analyseurs syntaxiques.

Dans ce contexte se dirige le développement d'un analyseur syntaxique pour le langage de spécification formelle Object Z. Object Z qui est une extension du langage de spécification Z est basé sur la théorie des ensembles, la logique des prédicats et l'orienté objet. Object Z utilise un langage schématique pour composer et structurer une série de descriptions mathématiques : il permet de grouper un ensemble d'informations (données ou opérations), de les encapsuler et les nommer pour réemploi.

Comme tout autre langage de spécification, il était vital d'avoir des outils pour la vérification et l'analyse de la syntaxe des spécifications écrites en Z et éventuellement en Object Z. La motivation principale qui a affirmé la nécessité de ces outils de vérification, était le nombre énorme de lignes de code qui peut le contenir une spécification d'un système informatique donné, et qui rend le fait de commettre des fautes, une chose inévitable, et rend la détection "manuelle" de ces fautes une tâche très difficile pour les experts du langage, et pire pour les débutants.

Il est donc intéressant de proposer un analyseur syntaxique permettant de vérifier les spécifications écrites en Object Z, d'identifier les conflits de spécifications (i.e., déclarations circulaires), d'afficher des messages d'erreur et de certifier leur correction. L'objectif central dans notre recherche, était la vérification de l'efficacité des algorithmes de compilation sur le langage Object Z, rappelant que cette efficacité est déjà montrée sur les langages de programmation. L'autre objectif était la combinaison des techniques de compilation avec des techniques permettant la résolution des contraintes sémantiques. Cette recherche a donné naissance à un nouvel analyseur syntaxique qu'on va le présenter dans cet article.

Cet article contient, dans la première section, une présentation de quelques systèmes de vérification Object Z existants. Nous effectuons ensuite une comparaison entre ces outils, en donnant les caractéristiques principales de chacun d'eux. Dans une autre section, nous recensons les contraintes sémantiques et les erreurs de types, qu'on les juge les plus fréquentes dans les spécifications. Finalement, nous proposons une démarche pour la construction d'un système d'analyse syntaxique des spécifications écrites en Object Z, basée sur notre expérience pratique dans ce domaine.

# 2. Aperçu sur l'Object Z

Object Z qui est une extension du langage Z a hérité tout sa syntaxe afin de faciliter la spécification dans un style orienté objet. L'avantage principal ajouté par Object Z était la notion de classe, qui a fourni à la spécification une grande flexibilité, et au spécificateur une facilité supplémentaire. La classe encapsule toutes les composantes (les constantes, les variables d'état, le schéma d'état initial et les opérations) dans une seule structure syntaxique, qui contient une interface (liste de visibilités) contrôlant les interactions entre les objets, qui sont des instances des classes, dans le même environnement.

```
┌─ NomClasse[PrametreGenerique] ──────────────────
│  Liste de visibilités
│  Classes héritées
│  Définitions locales
│  Schéma d'état
│  Schéma d'état initiale
│  Les opérations
└─────────────────────────────────────────────────
```

En Object Z, la portée d'un paramètre générique d'une classe générique est a classe entière. Par conséquent, à la différence de la spécification en Z, il n'y a aucun besoin de la présence du paramètre générique à chaque définition d'un schéma.

Object Z a introduit la notion d'héritage [2]. Une classe peut hériter les variables, les constantes, le schéma d'état initial et les opérations d'une autre classe, mais elle ne peut pas hériter sa liste de visibilité. La classe fille peut conserver ou modifier les composantes et la liste de visibilité de sa super-classe.

Cet aperçu va nous permettre à comprendre les contraintes sémantiques traitées lors de l'analyse sémantique.

## 3. Comparaison des systèmes existants

Dans ce paragraphe, nous présentons une étude comparative sur quelques outils de vérification des spécifications formelles écrites essentiellement en Z et Object Z. Cette étude montre d'une part, les avantages et les inconvénients de chacun des outils présentés et, d'autre part, va nous permettre de construire un analyseur syntaxique pour l'Object Z ultérieurement. Dans ce qui suit, nous allons traiter trois systèmes: Z-Eves, Z Animation System et Wizard.

### 3.1. Z-Eves

Z-Eves 2.1 [3] est un éditeur graphique qui inclut les fontes de Z (version 1.04) créer par Richard Jones [4], fonctionne sous Linux, Windows 95/98/NT/XP et Solaris. Il se divise en deux composants : une interface graphique et un serveur. Le premier composant gère les spécifications Z, transmet les commandes au serveur et affiche les résultats. Le deuxième vérifie les paragraphes et exécute les commandes de preuve. A l'entrée, Z-Eves importe les spécifications éditées par LaTeX dans des fichiers ayant le format .tex ou .zed. Notons que Z-Eves possède son propre format .zev. Z-Eves génère à la sortie des fichiers .zev, .rtf et .ps, l'exportation au format LaTeX n'est pas pris en charge dans la version 2.1 [5].

Le parser de Z-Eves est performant grâce à un module qui effectue les preuves d'erreurs. Aussi, son interface graphique présente une grande avantage, car elle fournie une grande facilité d'utilisation. Dans un autre coté, Z-eves présente quelques limites, comme l'incapacité de supporter le quantificateur d'unicité d'existence dans les expressions des schémas [3]. Aussi, une faiblesse remarquable au niveau de l'interopérabilité avec les autres outils de vérification (e.g. un nombre limité de format générés) et une incapacité de supporter les concepts orientés objet.

### 3.2. Z Animation System (ZANS)

Parmi les buts principaux de ZANS [6], est la validation des spécifications Z et le raffinement des conceptions basé sur le langage Z grâce a l'approche d'animation qui se base sur l'exécution des schémas opérationnels d'une spécification donnée, pour voir si elle répond exactement aux besoins ou non. La clé de cette approche est la classification des schémas opérationnels en deux catégories, un schéma explicite (toutes les variables de sortie de ce schéma sont définis par des variables d'entrée d'une façon directe ou indirecte) et un schéma non explicite (dite aussi implicite, dont certaines variables de sortie sont obligées à satisfaire à des contraintes construits par des variables d'entrées). Notons que seule les schémas explicites peuvent être exécutés (par conséquent animés) et que 94 % des schémas sont explicites ou qui peuvent être explicites avec des petites modifications [7]. Un autre objectif de ZANS est l'assistance de l'étude du langage Z, puisque il assure une composition facile des schémas et assure l'évaluation des prédicats. Ceci rend l'apprentissage de ce langage plus facile et plus interactif [8].

Actuellement, ZANS peut traiter les tâches suivantes : évaluation des expressions et des prédicats, vérification de type pour les spécifications Z, et exécution des schémas opérationnels, mais il ne peut pas animer les schémas non explicites. ZANS rencontre des difficultés dans la manipulation des ensembles infinis. Par exemple, les ensembles des entiers naturels et relatifs sont implémentés comme des ensembles finis ce qui impose des limites lors de l'utilisation de ces ensembles.

Le parseur de ZANS est pratiquement faible et inefficace, il est incomparable avec Z-EVES au niveau de la puissance. Ceci peut être expliqué par l'absence d'un module pour la preuve des théorèmes dans les spécifications. On peut aussi considérer l'absence d'une interface graphique un inconvénient qui diminue l'interactivité de cet outil. Encore, ZANS se limité au notation Latex (ou ZSL) au niveau des spécifications [8], ce qui impose aux utilisateurs de fournir des efforts supplémentaires.

### 3.3. Wizard

Wizard [9] est un vérificateur de types pour les spécifications écrites en Object Z. Ses tâches principales sont: la construction d'un arbre syntaxique abstrait, la vérification des noeuds de l'arbre et la génération des messages d'erreurs. Pour effectuer ces tâches, le système Wizard utilise les modules suivants : un analyseur lexical (scanner), un analyseur syntaxique (parseur), un générateur d'arbre abstrait, un générateur d'attributs et un vérificateur de types [10].

Wizard peut être considéré actuellement comme le plus puissant vérificateur des spécifications écrites en Object Z grâce à un parseur efficace et une grammaire bien développée qui couvre la plupart de la syntaxe d'Object Z. L'inconvénient principal de ce vérificateur est l'absence d'une interface graphique, puisque les commandes sont introduites par des commandes sur une boite DOS.

### 3.4. Evaluation des systèmes présentés et Récapitulation

Comme nous l'avons déjà remarqué, chaque outil de vérification admet sa propre technique pour l'analyse des spécifications Z ou Object Z, et ça n'interdit pas la présence de plusieurs points de ressemblances entre eux. Nous constatons qu'il y a un retard en nombre et en qualité des outils de vérifications du langage Object Z par rapport à celle du langage Z, et ce manque peut être expliqué par la complexité d'Object Z induite par ses concepts orienté objet. Dans le tableau de la figure 1, nous récapitulons les caractéristiques principales de quelques vérificateurs privilégiés, en se basant sur [5], [11], [12], [8] et [13].

|  | **Plate forme** | **Commandes** | **Input** | **Output** |
|---|---|---|---|---|
| **Z-Eves 2.1** | Linux, Windows, Solaris | Interface graphique | tex, zed, zev | RTF, ps, zev |
| **Piza 1.09** | Unix | Interface graphique, Prolog command prompt, Unix command prompt | pza | zev, z, zed, fuzz, tex, pza |
| **ZTC 2.03** | IBM PC, SUN | Ligne de commande DOS | zsl, zed | zsl, tex, zed, typ, log |
| **ZANS 0.3** | Win9x/Me/NT/XP, Unix Solaris, Sun, Linux | Ligne de commande DOS | zsl, zed | zsl, zed, log |
| **Fuzz 2.0** | Sun, IBM/PC, VAX/VMS | Ligne de commande | zed, tex, fuzz | fuzz, tex, zed |

**Tableau 1.** *Evaluation et synthèse*

Dans ce qui suit, nous identifions des contraintes sémantiques afin de l'intégrer ultérieurement dans l'analyseur syntaxique proposé.

### 4. Contraintes sémantiques

Nous présentons dans cette section quelques contraintes sémantiques que nous les tenons en considérations dans le développement de notre analyseur syntaxique. Ces contraintes sémantiques sont illustrées par des exemples en langage Z qui expliquent la nature et l'emplacement des fautes. Les fautes présentées sont essentiellement : la déclaration circulaire, l'existence d'un type non défini, une variable déclarée plusieurs fois dans un schéma, une variable qui porte un nom réservé pour un type, et la liste Δ contient une variable n'appartient pas à la liste des variables d'état.

### 4.1. Déclaration circulaire

La déclaration circulaire se produit quand on déclare une variable à partir d'une autre variable déjà déclarée dans le même schéma, comme l'indique le schéma *Exp1*.

```
┌─ Exp1 ──────────────
│ a : 𝔽 ℕ
│ b : a
└─────────────────────
```

Dans ce schéma, la variable *b* est déclarée à partir de la variable *a* déjà déclarée dans *Exp1*.

### 4.2. Type non défini

Le type non défini se produit quand on déclare une variable à partir d'un type non défini. A titre d'exemple, le type *Trame* dans le schéma *Exp2* n'est pas défini.

[Message]

```
┌─ Exp2 ──────────────
│ a : ℤ
│ b : Message
│ c : Trame
└─────────────────────
```

Notons que le type *Message* est un type de base.

### 4.3. Variable déclarée plusieurs fois

Cette faute se produit quand on déclare la même variable plusieurs fois dans le même schéma, comme l'indique le schéma *Exp3*.

```
┌─ Exp3 ──────────────
│ a : ℤ
│ a : ℕ
└─────────────────────
```

La variable *a* est déclarée deux fois dans le même schéma.

### 4.4. Variable portant un nom réservé pour un type

Cette faute se produit quand on déclare une variable portant le nom d'un type, comme le présente le schéma *Exp4*.

```
┌─ Exp4 ──────────────
│ Message : ℤ
└─────────────────────
```

On a déclaré *Message* comme un type. Donc on n'a pas le droit de le réutiliser comme une variable.

### 4.5. Liste Δ contient une variable n'appartenant pas aux variables d'état

Cette faute se produit si la liste Δ contient une variable qui n'appartient pas à l'ensemble des variables d'état. Le schéma *Exp5* illustre cette faute.

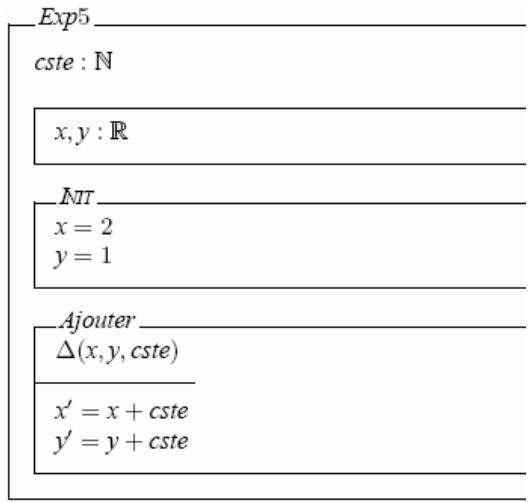

Dans la liste de paramètre Δ de l'opération *Ajouter*, on trouve la constante *cste* ce qui est considéré comme une erreur puisque *cste* n'est pas une variable d'état.

## 5. Démarche proposée

La démarche que nous proposons dans cette section va nous permettre de développer un analyseur syntaxique pour les spécifications écrites en Object Z. Pour ce faire, nous choisissons tout d'abord le format d'entrée. Ensuite, nous élaborons un système de règles. Finalement, nous implémentons les algorithmes nécessaires pour la construction du système.

### 5.1. Choix du format d'entrée

Nous avons suivi les outils, déjà décrits dans la section "Comparaison des systèmes existants". Ces outils utilisent les commandes LaTeX pour écrire les spécifications écrites en Z ou en object Z. Ce choix n'est pas aléatoire mais basé sur plusieurs points. En effet, la présence de deux fichiers de style zed.sty et oz.sty dans la bibliothèque de LaTeX, le premier développé par Mike Spivey et le second par Paul King (voir [14] et [15]) facilite l'écriture des spécifications. Ces fichiers contiennent des commandes nécessaires (et faciles) pour l'édition et l'impression des spécifications Z et Object Z. Un avantage est constaté. C'est que l'utilisation des commandes LaTeX va réduire d'une façon énorme les erreurs syntaxiques, puisque les règles dactylographiques de LaTeX vont être (en même temps) des règles syntaxiques pour les compilateurs. Pour ces raisons, et d'autres, nous éditons les fichiers d'entrées de notre compilateur avec LaTeX en utilisant les packages zed.sty et oz.sty.

### 5.2. Techniques d'analyse syntaxique utilisées

La grammaire proposée pour l'analyse syntaxique du langage Object-Z est de type ***SLR(1)***.

– ParagraphList = Paragraph
– ParagraphList = Paragraph ParagraphList
– Paragraph = \begin{class} { ClassHeading } \end{class}
– Paragraph = \begin{class} { ClassHeading } Visibility \inherit Inheritance \endinherit StateSchema
  InitialSchema Operations \end{class}

...

En fait, elle est inspirée de celle proposée dans [10]. Nous rappelons que l'analyseur syntaxique reçoit une suite d'unités lexicales (de symboles terminaux). Il doit dire si cette suite (ce mot) est syntaxiquement correct, c'est à dire si c'est un mot du langage généré par la grammaire qu'il possède. L'analyseur syntaxique construit l'arbre de dérivation de ce mot. S'il y arrive, alors le mot est syntaxiquement correct, sinon il est incorrect. Il existe deux méthodes pour construire cet arbre de dérivation : une méthode descendante et une méthode ascendante. Dans notre approche, nous utilisons la méthode ascendante. Ainsi, nous construisons un arbre de dérivation du bas (les feuilles, i.e. les unités lexicales) vers le haut (la racine, i.e. l'axiome de départ). Le modèle général utilisé est le modèle par décalages-réductions. Alors, on ne s'autorise que deux opérations :

– **décalage (shift)** : décaler d'une lettre le pointeur sur le mot en entrée.
– **réduction (reduce)** : réduire une chaîne (suite consécutive de terminaux et non terminaux à gauche du pointeur sur le mot en entrée et finissant sur ce pointeur) par un non-terminal en utilisant une des règles de production.

Ce modèle va permettre d'analyser les langages basés, surtout, sur les grammaires dites LR. Donc, nous l'utilisons pour analyser le langage Object Z en relation avec la grammaire proposée. Pour construire la table d'analyse de la grammaire proposée, on a besoin d'implémenter les algorithmes nécessaires pour le calcul des ensembles *PREMIER*, le calcul des ensembles *SUIVANT* et le calcul des Fermetures et des transitions des ensembles d'items avec leurs collections. Cette table d'analyse ainsi construite détermine si une spécification donnée *m* est générée ou non par la grammaire. Autrement, elle détermine si la spécification en question est syntaxiquement correcte ou non. Pour cela, une **Pile** est utilisée afin de faciliter le processus d'analyse. L'algorithme suivant décrit le fonctionnement de l'automate à pile face à un flux d'unités lexicales.

**Algorithme de l'analyseur syntaxique SLR**

Données : spécification *S*, table d'analyse *T*.
Initialisation de la pile : empiler *$*, empiler l'état *0*.
Un pointeur *ps* sur le premier mot de *S*. On note que les mots sont distingués par la présence des espaces.

Soit *X* le symbole en sommet de pile
Soit *a* le mot pointée par *ps*
Si *X* est un numéro d'état alors
    Si *T[X,a]*= décalage par l'état N° *n* alors
        Empiler *a* puis empiler *n*
        Avancer *ps*
    Si *T[X,a]*= réduction par la règle *y* → *Y₁ Y₂...Yₙ* alors
        Dépiler $Y_n$, $Y_{n-1}$, ..., $Y_1$ et dépiler en parallèle tout les états intermédiaire
        Empiler *y*
    Si *a = $* et *T[X,a]= ACC* alors *ACCEPTER*
    Sinon *ERREUR*
    FinSi
Sinon Si *X* est une non terminale alors
    Soit *y* le symbole juste avant le *X* dans la pile
    Si *y* est un numéro d'état alors
        Si *T[y,X]*= numéro d'un état soit *n* alors empiler *n*
        Sinon *ERREUR*
        FinSi
    Sinon *ERREUR*
    FinSi
Sinon *ERREUR*
FinSi
Jusqu'à *ERREUR* ou *ACCEPTER*

En se basant sur l'algorithme ci-dessus, nous analysons l'exemple de la spécification simple suivante. Cet exemple présente une classe vide nommée *A* :

Le code en LaTeX de cette classe est comme suit :

```
\begin{verbatim}
\begin{class} { A }
\end{class}
\end{verbatim}
```

L'état de la pile et les actions effectuées au cours de l'analyse syntaxique est présenté dans la figure 1. Notons que les états sont distingués dans la figure par un carré.

| Pile | entrée | action |
|---|---|---|
| $ [0] | \begin{class} { A } \end{class} $ | d1 |
| $ [0] \begin{class} [1] | { A } \end{class} $ | d5 |
| $ [0] \begin{class} [1] { [5] | A } \end{class} $ | d2 |
| $ [0] \begin{class} [1] { [5] A [2] | } \end{class} $ | r4: ClassHeading = Word |
| $ [0] \begin{class} [1] { [5] ClassHeading | } \end{class} $ | Je suis en 5 avec ClassHeading : je vais en 7 |
| $ [0] \begin{class} [1] { [5] ClassHeading [7] | } \end{class} $ | d8 |
| $ [0] \begin{class} [1] { [5] ClassHeading [7] } [8] | \end{class} $ | d9 |
| $ [0] \begin{class} [1] { [5] ClassHeading [7] } [8] \end{class} [9] | $ | r3: Paragraph = \begin{class} { ClassHeading } \end{class} |
| $ [0] Paragraph | $ | Je suis en 0 avec Paragraph : je vais en 4 |
| $ [0] Paragraph [4] | $ | r1 : ParagraphList = Paragraph |
| $ [0] ParagraphList | $ | Je suis en 0 avec ParagraphList : je vais en 3 |
| $ [0] ParagraphList [3] | $ | ACCEPTE |

**Figure 1.** *Exemple d'analyse d'une spécification*

Les algorithmes ainsi développés possèdent des complexités acceptables. Notons que la grammaire utilisée par l'analyseur contient environs 124 règles ce qui nous donne une table d'analyse de taille 250×94.

## 6. Expérimentation et présentation du prototype

Dans cette section, nous présentons tout d'abord le prototype développé. Ensuite, nous étudions son comportement face aux erreurs syntaxiques et sémantiques.

L'analyseur syntaxique développé prend à l'entrée une spécification écrite en Object Z éditée avec le code LaTeX. Il effectue deux types de vérifications : syntaxiques et sémantiques en générant, si nécessaire, des messages d'erreurs. Le code cible de cet analyseur est aussi LaTeX.

Le programme représentant le compilateur est écrit en langage Java. Le choix de Java est justifié par la puissance et la flexibilité de ce langage orienté objet qui est un langage fortement typé.

Le développement de l'analyseur syntaxique a passé par trois niveaux. Le premier niveau est l'implémentation du noyau du programme qui est composé par les différents algorithmes nécessaires pour la manipulation des grammaires et l'analyse des spécifications. Le deuxième niveau est la conception et la réalisation d'une interface graphique interactive et adéquate au fonctionnement du programme. Le troisième niveau est l'embarquement de l'interface graphique sur le noyau du programme.

L'exemple de la spécification d'une file d'attente générique ***Queue[Item]*** extrait de [2] est utilisé pour illustrer le processus d'analyse. Notons qu'il faut séparés les différentes unités lexicales de la spécification par des blancs (des espaces) pour qu'elles puissent être reconnues par le parseur.

```
\begin{class} { Queue [ Item ] }
\visibility ( count , Init , Join , Leave )
\begin{state}
items : \seq  Item \\
count : \nat
\end{state}
\begin{init}
items = \emptyseq \\
count = 0
\end{init}
\begin{op} { Join }
\Delta ( items , count )
item? : Item
\ST
items' = items \cat \lseq
 item? \rseq \\
count' = count + 1
\end{op}
\begin{op} { Leave }
\Delta ( items )
item! : Item
\ST
items = \lseq item!
\rseq \cat items'
\end{op}
\end{class}
```

Supposons que nous avons une erreur syntaxique dans la partie déclarative du schéma d'état (voir figure 2). Par exemple, les deux points de déclaration (:) sont remplacés par l'opérateur d'égalité (=), le nouveau schéma d'état devient :

```
\begin{state}
items : \seq  Item \\
count = \nat
\end{state}
```

Comme l'indique la figure 2, l'analyseur va détecter l'erreur et affiche le message suivant : *Classe "Queue", une erreur dans le schéma d'état : le syntaxe est incorrecte et ceci est causé par la chaîne "=".*

L'analyseur syntaxique peut localiser l'erreur à plusieurs niveaux. Le premier niveau est le nom de la classe où se produit l'erreur, cela est utile si la spécification contient plusieurs classes. Le deuxième niveau est le nom du bloc contenant l'erreur : qui peut être l'entête de la classe, la liste de visibilité, la liste d'héritage, le schéma d'état, le schéma d'état initial ou les opérations. Le troisième et le dernier niveau dans l'opération de localisation de l'erreur est le nom du symbole qui a causé l'erreur (i.e., le symbole qui a causé l'arrêt de l'analyse syntaxique). Ainsi, en connaissant la classe, le bloc ou le symbole responsable de l'erreur, l'utilisateur peut borner facilement l'erreur syntaxique et par suite peut la corriger.

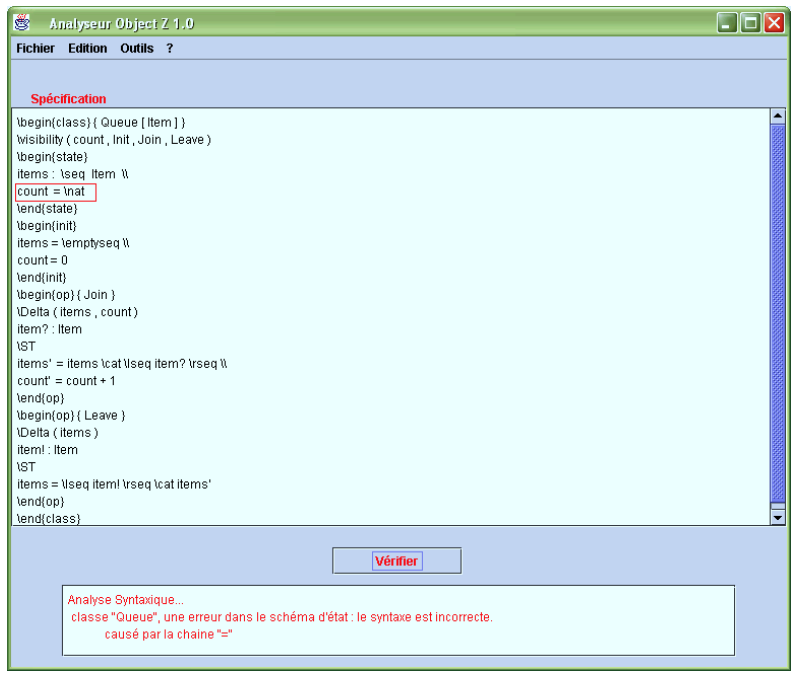

**Figure 2.** *Erreur syntaxique*

Après avoir expérimenté le prototype développé sur des erreurs syntaxiques, nous testons maintenant son comportement face aux contraintes sémantiques déjà présentées dans la section "contraintes sémantiques". Nous reprendrons la spécification de la file d'attente générique **_Queue[Item]_** et nous provoquons deux erreurs sémantiques en même temps : la première est une déclaration circulaire et la deuxième est un type non défini.

[ Message ]
\begin{class} { Queue [ Item ] }
\visibility ( count , Init , Join , Leave )
\begin{state}
items : \seq  Item \\
count : \nat \\
mess : \pset Message \\
m : mess
\end{state}
\begin{init}
items = \emptyseq \\
count = 0
\end{init}
\begin{op} { Join }
\Delta ( items , count )
item? : Item
\ST
items' = items \cat \lseq
item? \rseq \\
count' = count + 1
\end{op}
\begin{op} { Leave }
\Delta ( items )
item! : Item
\ST

```
items = \lseq item! \rseq
\cat items'
\end{op}
\end{class}
```

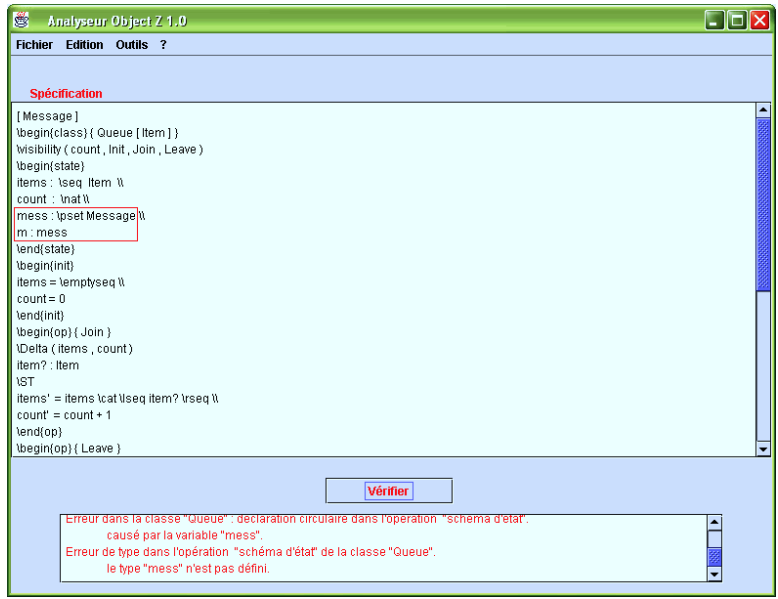

**Figure 3.** *Erreurs sémantiques*

Comme l'indique la partie encadrée dans la zone de spécification de la figure 3, on déclare un nouveau type *Message* et on ajoute au schéma d'état les instructions suivantes :

```
mess : \pset Message \\
m : mess
```

Ainsi, on a déclaré la variable *mess* à partir du type *Message* et la variable *m* à partir du variable *mess*. Suite à la vérification de cette spécification, l'analyseur va détecter deux erreurs sémantiques et il affiche le message suivant : *Erreur dans la classe "Queue" : déclaration circulaire dans l'opération "schéma d'état" causée par la variable "mes". Erreur de type dans l'opération "schéma d'état" de la classe "Queue". Le type "mess" n'est pas défini.*

## 7. Conclusion et discussion

Dans cet article, nous avons fourni une démarche pour la construction d'un analyseur syntaxique et sémantique pour les spécifications écrites en Object Z. Tout d'abord, nous avons réalisé une comparaison entre quelques outils de vérification afin de prendre une idée sur l'état de l'art actuel, en fournissant un tableau récapitulatif qui recense pour chaque outil ses caractéristiques principales. Ensuite, nous avons présenté quelques contraintes sémantiques en Object Z, illustrées par des exemples. Puis, nous avons donné une idée sur la technique de compilation utilisée. Finalement, nous avons présenté un exemple qui montre l'efficacité de l'analyseur développé face aux erreurs syntaxiques et sémantiques. Actuellement, nous sommes en train d'améliorer la performance de l'analyseur syntaxique. En effet, nous voulons, d'une part, intégrer des techniques d'optimisation dans l'analyseur syntaxique, et d'autre part, de l'enrichir à fin d'écrire un compilateur du langage Object Z vers un langage de plus bas niveau à plus long terme.

Bibliographie